\documentclass[a4paper,UKenglish,cleveref, autoref, thm-restate]{lipics-v2019}

\pdfoutput=1

\title{Idris 2: Quantitative Type Theory in Practice} %TODO Please add

\author{Edwin Brady}
{School of Computer Science, University of St Andrews, Scotland, UK \and \url{http://www.type-driven.org.uk/edwinb}}
{ecb10@st-andrews.ac.uk}{https://orcid.org/0000-0002-9734-367X}
{}%TODO mandatory, please use full name; only 1 author per \author macro; first two parameters are mandatory, other parameters can be empty. Please provide at least the name of the affiliation and the country. The full address is optional

\authorrunning{E. Brady} %TODO mandatory. First: Use abbreviated first/middle names. Second (only in severe cases): Use first author plus 'et al.'

\Copyright{Edwin Brady} %TODO mandatory, please use full first names. LIPIcs license is "CC-BY";  http://creativecommons.org/licenses/by/3.0/

\ccsdesc[500]{Software and its engineering~Functional languages}

\keywords{Dependent types, linear types, concurrency} %TODO mandatory; please add comma-separated list of keywords

\category{} %optional, e.g. invited paper

\relatedversion{} %optional, e.g. full version hosted on arXiv, HAL, or other respository/website
\supplement{}%optional, e.g. related research data, source code, ... hosted on a repository like zenodo, figshare, GitHub, ...

\nolinenumbers %uncomment to disable line numbering

\hideLIPIcs  %uncomment to remove references to LIPIcs series (logo, DOI, ...), e.g. when preparing a pre-final version to be uploaded to arXiv or another public repository

\EventEditors{} % John Q. Open and Joan R. Access}
\EventNoEds{2}
\EventLongTitle{ECOOP 2021} % 42nd Conference on Very Important Topics (CVIT 2016)}
\EventShortTitle{ECOOP 2021}
\EventAcronym{ECOOP}
\EventYear{2021}
\EventDate{December 24--27, 2016}
\EventLocation{Little Whinging, United Kingdom}
\EventLogo{}
\SeriesVolume{42}
\ArticleNo{23}
\usepackage{idrislang}
\begin{document}

\maketitle

\begin{abstract}
% context
    Dependent types allow us to express precisely \emph{what} a function is
    intended to do.  Recent work on Quantitative Type Theory (QTT) extends
    dependent type systems with \emph{linearity}, also allowing precision in
    expressing \emph{when} a function can run.
% gap
    This is promising, because it suggests the ability to design and
    reason about resource usage protocols, such as we might find in
    distributed and concurrent programming, where the state of a communication
    channel changes throughout program execution.
    As yet, however, there has not been a full-scale programming language with
    which to experiment with these ideas.
% innovation
    Idris 2 is a new version of the dependently typed language Idris, with a
    new core language based on QTT, supporting linear and dependent types.  In
    this paper, we introduce Idris 2, and describe how QTT has influenced its
    design. We give examples of the benefits of QTT in practice including:
    expressing which data is erased at run time, at the type level; and,
    resource tracking in the type system leading to type-safe concurrent
    programming with session types.
\end{abstract}

\section{Introduction}

Dependently typed programming languages, such as Idris~\cite{idris-jfp},
Agda~\cite{norell2007thesis}, and Haskell with the appropriate extensions 
enabled~\cite{dephaskell}, allow us to give precise types which can
describe assumptions about and relationships between inputs and outputs.
This is valuable for reasoning about \emph{functional} properties, such as
correctness of algorithms on collections~\cite{McBride2014}, termination of
parsing~\cite{Danielsson2010} and 
scope safety of programs~\cite{allais-cpp17}.
However, reasoning about \emph{non-functional} properties in this setting,
such as memory safety, protocol correctness, or resource safety in general, is
more difficult though it can be achieved with techniques such as embedded
domain specific languages~\cite{brady-tfp14} or indexed
monads~\cite{Atkey2009,McBride2011}. These are, nevertheless, heavyweight
techniques which can be hard to compose.

Substructural type systems, such as linear type
systems~\cite{wadler-linear,Morris2016,linhaskell},
allow us to express \emph{when} an operation can be executed, by requiring
that a linear resource be accessed \emph{exactly once}. Being able to
combine linear and dependent types would give us the ability to express
an ordering on operations, as required by a protocol, with precision
on exactly what operations are allowed, at what time. Historically, however,
a difficulty in combining linear and dependent types has been in deciding
how to treat occurrences of variables in \emph{types}. This can be
avoided~\cite{Krishnaswami-ldtp} by never allowing types to depend on a
linear term, but more recent work on Quantitative Type Theory 
(QTT)~\cite{qtt,nuttin} solves the problem by assigning a \emph{quantity} to
each binder, and checking terms at a specific \emph{multiplicity}. 
Informally, in QTT, variables and function arguments have a multiplicity: $0$,
$1$ or unrestricted ($\omega$). We can freely use any variable in an argument
with multiplicity $0$---e.g., in types---but we can not use a variable with
multiplicity $0$ in an argument with any other
multiplicity. A variable bound with multiplicity $1$ must be used exactly once.
In this way, we can describe linear resource usage protocols, and furthermore
clearly express \emph{erasure} properties in types.
%We will make the concept
%of multiplicities more precise later in the paper.

Idris 2 is a new implementation of Idris, which uses QTT as its core
type theory. In this paper, we explore the possibilities of programming with
a full-scale language built on QTT. By full-scale, we mean a language with high
level features such as inference, interfaces, local function definitions
and other syntactic sugar. As an example, we will show how to implement a
library for concurrent programming with session types~\cite{Honda1993}.
We choose this example because, as demonstrated by the extensive literature
on the topic, correct concurrent programming is both hard to achieve, and
vital to the success of modern software engineering. Our aim is to show that
a language based on QTT is an ideal environment in
which to implement accessible tools for software developers, based on
decades of theoretical results.

%The code is
%submitted as anonymised supplementary material (\texttt{idris2-code.tgz}).

\subsection{Contributions}

This paper is about exploring what is possible in a language based on
Quantitative Type Theory (QTT), and introduces a new implementation of Idris.
We make the following research contributions:

\begin{itemize}
\item We describe Idris 2 (Section~\ref{sec:idris}), the first implementation
of quantitative type theory in a full programming language, and the first
language with full first-class dependent types implemented in itself.
\item We show how Idris 2 supports two important applications of quantities:
\emph{erasure} (Section~\ref{sec:erasure}) which gives type-level guarantees as
to which values are required at run-time, and \emph{linearity}
(Section~\ref{sec:linearity}) which gives type-level guarantees of resource
usage. We also describe a general approach to implementing linear
resource usage protocols (Section~\ref{sec:protocols}).
\item We give an example of QTT in practice, encoding bidirectional
session types (Section~\ref{sec:sessions}) for safe concurrent programming.
\end{itemize}

We do not discuss the metatheory of QTT, nor the trade-offs in its design
in any detail. Instead, our interest is in discussing how it has affected
the design of Idris 2, and in investigating the new kinds of programming and
reasoning it enables. Where appropriate, we will discuss the intuition behind
how argument multiplicities work in practice.

\section{An Overview of Idris}

\label{sec:idris}

Idris is a purely functional programming language, with \emph{first-class}
dependent types. That is, types can be computed, passed to and returned from
functions, and stored in variables, just like any other value. In this section,
we give a brief overview of the fundamental features which we 
use in this paper. A full tutorial is available
online\footnote{\url{https://idris2.readthedocs.io/en/latest/tutorial/index.html}}.
Readers who are already familiar with Idris may skip to Section~\ref{sec:idris2}
which introduces the new implementation.

\subsection{Functions and Data Types}

\label{sec:typevars}

The syntax of Idris is heavily influenced by the syntax of Haskell. Function
application is by juxtaposition and, like Haskell and ML and other related
languages, functions are defined by recursive pattern matching equations. For
example, to append two lists:

\begin{lstlisting}
append : List a -> List a -> List a
append [] ys = ys
append (x :: xs) ys = x :: append xs ys
\end{lstlisting}

The first line is a \demph{type declaration}, which is required in
Idris\footnote{Note that unlike Haskell, we use a single colon for the type
declaration.}. Names in the type declaration which begin with a lower case
letter are \emph{type level variables}, therefore \TC{append} is parameterised
by an element type.
Data types, like \TC{List}, are defined in terms of their \demph{type
constructor} and \demph{data constructors}:

\begin{lstlisting}
data List : Type -> Type where
     Nil  : List elem
     (::) : elem -> List elem -> List elem
\end{lstlisting}

The type of types is \TC{Type}.
%\footnote{There is also an infinite hierarchy
%of cumulative universes, but the details are beyond the scope of this paper}
Therefore, this declaration states that \TC{List} is parameterised by a
\TC{Type}, and can be constructed using either \TC{Nil} (an empty list) or
\TC{::} (``cons'', a list consisting of a head element and and tail).
As we'll see in more detail shortly, types in Idris are first-class, thus
the type of \TC{List} (\TC{Type -> Type}) is an ordinary function type.
Syntax sugar allows us to write \texttt{[]} for \DC{Nil}, and comma separated
values in brackets expand to applications of \texttt{::}, e.g. \texttt{[1, 2]}
expands to \texttt{1 :: 2 :: Nil}.

\subsection{Interactive Programs}

\label{sec:interactive}

Idris is a pure language, therefore functions have no side effects. Like
Haskell~\cite{awkward}, we write interactive programs by \emph{describing} interactions using
a parameterised type \TC{IO}. For example, we have primitives for Console I/O, including:

\begin{lstlisting}
putStrLn : String -> IO ()
getLine : IO String
\end{lstlisting}

\TC{IO t} is the type of an interactive action which produces a result of
type \TC{t}. So, \TC{getLine} is an interactive action which, when executed,
produces a \TC{String} read from the console. Idris the language \emph{evaluates}
expressions to produce a description of an interactive action as an \TC{IO t}.
It is the job of the run time system to \emph{execute} the resulting action.
Actions can be chained using the $\mathtt{>>=}$ operator:

\begin{lstlisting}
(>>=) : IO a -> (a -> IO b) -> IO b
\end{lstlisting}

For example, to read a line from the console then echo the input:

\begin{lstlisting}
echo : IO ()
echo = getLine >>= \x => 
       putStrLn x
\end{lstlisting}

For readability, again like Haskell, Idris provides \texttt{do}-notation which
translates an imperative style syntax to applications of $\mathtt{>>=}$. The
following definition is equivalent to the definition of \texttt{echo}
above.

\begin{lstlisting}
echo : IO ()
echo = do x <- getLine
          putStrLn x
\end{lstlisting}

The translation from \texttt{do}-notation to applications of $\mathtt{>>=}$
is purely syntactic. In practice therefore we can, and do, use a more general
type for $\mathtt{>>=}$ to use \texttt{do}-notation in other contexts. We will
see this later, when implementing linear resource usage protocols.

\subsection{First-Class Types}

The main distinguishing feature of Idris compared to other languages, even
some other languages with dependent types, is that types are \emph{first-class}.
This means that we can use types anywhere that we can use any other value.
For example we can pass them as arguments to functions, return them from
functions, or store them in variables. This enables us to define type synonyms,
to compute types from data, and express relationships between and properties
of data.
% 
% \subsubsection{Type Synonyms}
% 
As an initial example, we can define a \emph{type synonym} as follows:

\begin{lstlisting}
Point : Type
Point = (Int, Int)
\end{lstlisting}

Wherever the type checker sees \FN{Point} it will evaluate it,
and treat it as \TC{(Int, Int)}:

\begin{lstlisting}
moveRight : Point -> Point
moveRight (x, y) = (x + 1, y)
\end{lstlisting}

Languages often include type synonyms as a special feature (e.g.
\texttt{typedef} in C or \texttt{type} declarations in Haskell). In Idris,
no special feature is needed. 
%First-class types also allow us to define: \emph{variadic functions}, where
%an argument is used to compute the rest of a function's type; and
%\emph{dependent data types}, where types are parameterised by values.

\subsubsection{Computing Types}

\label{sec:printf}

First-class types allow us to compute types from data.
A well-known example is \texttt{printf} in C, where
a format string determines the types of arguments to be printed.
C compilers typically use extensions to check the validity of the format string;
first-class types allow us to implement a \texttt{printf}-style variadic
function, with compile time checking of the format. We begin by defining
valid formats (limited to numbers, strings, and literals here):

\begin{lstlisting}
data Format : Type where
     Num : Format -> Format
     Str : Format -> Format
     Lit : String -> Format -> Format
     End : Format
\end{lstlisting}

This describes the expected input types. We can calculate
a corresponding function type:
  
\begin{lstlisting}
PrintfType : Format -> Type
PrintfType (Num f) = (i : Int) -> PrintfType f
PrintfType (Str f) = (str : String) -> PrintfType f
PrintfType (Lit str f) = PrintfType f
PrintfType End = String
\end{lstlisting}

A function which computes a type can be used anywhere that Idris expects a
value of type \TC{Type}. So, for the type of \FN{printf}, we name the first
argument \VV{fmt}, and use it to compute the rest of the type of \FN{printf}:

\begin{lstlisting}
printf : (fmt : Format) -> PrintfType fmt
\end{lstlisting}

We can check the type of an expression, even using an as yet undefined function,
at the Idris REPL. For example, a format corresponding to \texttt{"\%d \%s"}:

\begin{lstlisting}
Main> :t printf (Num (Lit " " (Str End)))
printf (Num (Lit " " (Str End))) : Int -> String -> String
\end{lstlisting}

We will implement this via a helper function which accumulates a string:

\begin{lstlisting}
printfFmt : (fmt : Format) -> (acc : String) -> PrintfType fmt
\end{lstlisting}

Idris has support
for \emph{interactive} program development, via text editor plugins and REPL
commands. In this paper, we will use \emph{holes} extensively.
An expression of the form \texttt{?hole} stands for an as yet unimplemented
part of a program. This defines a top level function \texttt{hole}, with a type
but no definition, which we can inspect at the REPL. So, we can write a partial
definition of \texttt{printfFmt}:

\begin{lstlisting}
printfFmt : (fmt : Format) -> (acc : String) -> PrintfType fmt
printfFmt (Num f) acc = ?printfFmt_rhs_1
printfFmt (Str f) acc = ?printfFmt_rhs_2
printfFmt (Lit str f) acc = ?printfFmt_rhs_3
printfFmt End acc = ?printfFmt_rhs_4
\end{lstlisting}

Then, if we inspect the type of \texttt{printfFmt\_rhs\_1} at the REPL, we
can see the types of the variables in scope, and the expected type of
the right hand side:

\begin{lstlisting}
Example> :t printfFmt_rhs_1
   f : Format
   acc : String
------------------------------
printfFmt_rhs_1 : Int -> PrintfType f
\end{lstlisting}

So, a format specifier of \DC{Num} means we need to write a function that
expects an \TC{Int}. 
%
% Similarly, if we inspect the type of
% \texttt{printfFmt\_rhs\_2}, we'll see that a format specifier of \DC{Str} means
% we need to write a function that expects a \TC{String}:
% 
% \begin{lstlisting}
%    f : Format
%    acc : String
% ------------------------------
% printfFmt_rhs_2 : String -> PrintfType f
% \end{lstlisting}
%
For reference, the complete definition is given in Listing~\ref{printfdef}.
As a final step (omitted here) we can write a C-style \texttt{printf} by
converting a \TC{String} to a \TC{Format} specifier.

\begin{lstlisting}[caption={The complete definition of \texttt{printf}},label=printfdef,float=h]
printfFmt : (fmt : Format) -> (acc : String) -> PrintfType fmt
printfFmt (Num f) acc = \i => printfFmt f (acc ++ show i)
printfFmt (Str f) acc = \str => printfFmt f (acc ++ str)
printfFmt (Lit str f) acc = printfFmt f (acc ++ str)
printfFmt End acc = acc
  
printf : (fmt : Format) -> PrintfType fmt
printf fmt = printfFmt fmt ""
\end{lstlisting}

This example uses data which is known at compile time---the format
specifier---to calculate the rest of the type. In Section~\ref{sec:protocols},
we will see a similar idea used to calculate a type from data which is not
known until \emph{run time}.

\subsubsection{Dependent Data Types}

We define data types (such as \TC{List a} and \TC{Format} earlier) by giving
explicit types for the \emph{type constructor} and the \emph{data
constructors}.
We defined \TC{List} as having type \TC{Type -> Type}, a function type, since
\TC{List} is parameterised by an element type. We are
not limited to parameterising by types;
type constructors can be parameterised by any value. The canonical example
is a vector, \TC{Vect}, a list with its length encoded in the type:

\begin{lstlisting}
data Vect : Nat -> Type -> Type where
     Nil  : Vect Z a
     (::) : a -> Vect k a -> Vect (S k) a
\end{lstlisting}

\TC{Nat} is the type of natural numbers, declared as follows as part of the
Idris Prelude (its standard library), where \texttt{Z} stands for ``zero'' and
\texttt{S} stands for ``successor'', here using a Haskell-style \texttt{data}
declaration, which is sugar for the fully explicit type and data constructors:

\begin{lstlisting}
data Nat = Z | S Nat
\end{lstlisting}

\textbf{Note:}
%It is convenient to use this unary declaration of numbers to
%describe lengths and sizes of other structures, because then the structure of
%\texttt{Nat} allows a \texttt{S} to correspond to adding an element. 
Idris
optimises \texttt{Nat} to a binary representation at run time, although
in practice, values of type \texttt{Nat} can often be erased when the
type system can guarantee they are not needed at run time.

Since vectors have their lengths encoded in their types, functions over vectors
encode their effect on vector length in their types:

\begin{lstlisting}
append : Vect n a -> Vect m a -> Vect (n + m) a
append [] ys = ys
append (x :: xs) ys = x :: append xs ys
\end{lstlisting}

%We will see several examples of dependent data types throughout this paper.

\subsection{Implicit Arguments}

\label{sec:implicits}

As we noted in Section~\ref{sec:typevars}, lower case names in type definitions
are type level variables. So, in the following definition\ldots

\begin{lstlisting}
append : Vect n a -> Vect m a -> Vect (n + m) a
\end{lstlisting}

\ldots \VV{n}, \VV{a} and \VV{m} are type level variables. Note that we do not
say ``type variable'' since they are not necessarily of type \TC{Type}!
Type level variables are bound as \emph{implicit arguments}. Written
out in full, the type of \FN{append} is:

\begin{lstlisting}
append : {n : Nat} -> {m : Nat} -> {a : Type} ->
         Vect n a -> Vect m a -> Vect (n + m) a
\end{lstlisting}

\textbf{Note:} We will refine this when introducing multiplicities in
Section~\ref{sec:quantities}.

The implicit arguments are concrete arguments to \FN{append}, like the
\TC{Vect} arguments. Their values are solved by
unification~\cite{millerunification,gundrythesis}.  Typically implicit
arguments are only necessary at compile time, and unused at run time.
However, we can use an implicit argument in a definition, since
the names of the arguments are in scope in the right hand side:

\begin{lstlisting}
length : {n : Nat} -> Vect n a -> Nat
length xs = n
\end{lstlisting}

One challenge with first-class types is in distinguishing those parts of
programs which are required at run time, and those which can be erased.
Tradtionally, this phase distinction has been clear: types are erased at run
time, values preserved. But this correspondence is merely a coincidence,
arising from the special (non first-class) status of types! As we see with
\FN{length} and \FN{append}, sometimes an argument might be required (in
\FN{length}) and sometimes it might be erasable (in \FN{append}).
Idris 1 uses a constraint solving algorithm~\cite{matusicfp}, which has been
effective in practice, but has a weakness that it is not possible to tell from
a definition's type alone which arguments are required at run time.  In
Section~\ref{sec:erasure} we will see how quantitative type theory allows us to
make a precise distinction between the run time relevant parts of a program and
the compile time only parts.

\subsection{Idris 2}

\label{sec:idris2}

Idris 2 is a new version of Idris, implemented in itself, and based on
Quantitative Type Theory (QTT) as defined in recent work by Atkey~\cite{qtt}
following initial ideas by McBride~\cite{nuttin}. 
In QTT, each variable binding is associated with a \emph{quantity} (or
\emph{multiplicity}) which denotes the number of times a variable can be
used in its scope: either zero, exactly once, or unrestricted. We will
describe these further shortly. 
Several factors have motivated the new implementation:

\begin{itemize}
\item In implementing Idris in itself, we have necessarily done the engineering
required on Idris to implement a system of this scale. Furthermore, although it is
outside the scope of the present paper, we can explore the benefits of
dependently typed programming in implementing a full-scale programming language.
\item A limitation of Idris 1 is that it is not always clear which arguments to
functions and constructors are required \emph{at run time}, and which are 
erased, even despite previous work~\cite{Brady2005,matusicfp,matus-draft}. QTT allows us
to state clearly, in a type, which arguments are erased.
Erased arguments are still relevant \emph{at compile time}.
\item There has, up to now, been no full-scale implementation of a language
based on QTT which allows exploration of the possibilities of linear and
dependent types.
\item Purely pragmatically, Idris 1 has outgrown the requirements of its
initial experimental implementation, and since significant re-engineering has
been required, it was felt that it was time to re-implement it in Idris itself.
\end{itemize}

In the following sections, we will discuss the new features in Idris 2 which
arise as a result of using QTT as the core: firstly, how to express
\emph{erasure} in the type system; and secondly, how to encode resource usage
protocols using linearity.

\section{Quantities in Types}

\label{sec:quantities}

The biggest distinction between Idris 2 and its predecessor is that it is
built on Quantitative Type Theory (QTT)~\cite{nuttin,qtt}. In QTT, each
variable binding (including function arguments) has a \emph{multiplicity}.
QTT itself takes a general approach to multiplicities, allowing any semiring.
For Idris 2, we make a concrete choice of semiring, where a
multiplicity can be one of:

\begin{itemize}
\item $0$: the variable is not used at run time
\item $1$: the variable is used exactly once at run time
\item $\omega$: no restrictions on the variable's usage at run time
\end{itemize}

In this section, we describe the syntax and informal semantics of
multiplicities, and discuss the most important practical applications:
erasure, and linearity. The formal semantics are presented
elsewhere~\cite{qtt}; here, we aim to describe the intuition.
In summary:

\begin{itemize}
\item Variable bindings have multiplicities which describe how often the
varible must be used within the scope of its binding.
\item A variable is ``used'' when it appears in the body of a definition (that
is, not a type declaration), in an argument position with multiplicity $1$ or
$\omega$.
\item A function's type gives the multiplicities the arguments have in the
function's body.
\end{itemize}

Variables with multiplicity $\omega$ are truly unrestricted, meaning that they
can be passed to argument positions with multiplicity $0$, $1$ or $\omega$.
A function which
takes an argument with multiplicity $1$ promises that it will not share the
argument in the future; there is no requirement that it has not been shared
in the past.

\subsection{Syntax}

When declaring a funcion type we can, optionally, give an explicit multiplicity
of \texttt{0} or \texttt{1}. If an argument has no multiplicity
given, it defaults to $\omega$.  For example, we can declare the type of
an identity function which takes its polymorpic type \emph{explicitly}, and
mark it erased:

\begin{lstlisting}
id_explicit : (0 a : Type) -> a -> a 
\end{lstlisting}

If we give a partial definition of \FN{id\_explicit}, then inspect the type of
the hole in the right hand side, we can see the multiplicities of the variables
in scope:

\begin{lstlisting}
id_explicit a x = ?id_explicit_rhs

Main> :t id_explicit_rhs
0 a : Type
  x : a
------------------------------
id_explicit_rhs : a
\end{lstlisting}

This means that \VV{a} is not available at run time, and \VV{x} is unrestricted
at run time. If there is no explicit multiplicity given, it is $\omega$. A
variable which is not available at run time can only be used in
argument positions with multiplicity \texttt{0}.
Implicitly bound arguments are also given
multiplicity \texttt{0}. So, in the following declaration of \FN{append}\ldots

\begin{lstlisting}
append : Vect n a -> Vect m a -> Vect (n + m) a
\end{lstlisting}

\ldots \FN{n}, \FN{a} and \FN{m} have multiplicity \texttt{0}. In Idris 2,
therefore, unlike Idris 1, the declaration is equivalent to writing:

\begin{lstlisting}
append : {0 n : Nat} -> {0 m : Nat} -> {0 a : Type} ->
         Vect n a -> Vect m a -> Vect (n + m) a
\end{lstlisting}

The default multiplicities for arguments are, therefore:

\begin{itemize}
\item If you explicitly write the binding, the default is $\omega$.
\item If you omit the binding (e.g. in a type level variable), the default is $0$.
\end{itemize}

This design choice, that type level variables are by default
erased, is primarily influenced by the common usage in Idris 1, that
implicit type level variables are typically compile-time only.
As a result, the most common use cases involve the most lightweight syntax.

\subsection{Erasure}

\label{sec:erasure}

The multiplicity \texttt{0} gives us control over whether a function
argument---\TC{Type} or otherwise---is used at run time. This is important in
a language with first-class types since we often parameterise types by values
in order to make relationships between data explicit, or to make assumptions
explicit. In this section, we will consider two examples of this, and see
how multiplicity 0 allows us to control which data is available at run time.

\subsubsection{Example 1: Vector length}

We have seen how a vector's type includes its length;
we can use this length at run time, even though it is
part of the type, provided that it has non-zero multiplicity:

\begin{lstlisting}
length : {n : Nat} -> Vect n a -> Nat
length xs = n
\end{lstlisting}

With this definition, the length \VV{n} is available at run time, which has
a storage cost, as well as potentially the cost of computing the length
elsewhere. If we want the length to be erased, we would need to recompute it
in the \FN{length} function:

\begin{lstlisting}
length : Vect n a -> Nat
length [] = Z
length (_ :: xs) = S (length xs)
\end{lstlisting}

The type of \FN{length} in each case makes it explicit whether or not the
length of the \TC{Vect} is available. Let us now consider a more realistic
example, using the type system to ensure soundness of a compressed encoding
of a list.

\subsubsection{Example 2: Run-length Encoding of Lists}

Run-length encoding is a compression scheme which collapses sequences (runs)
of the same element in a list. It was originally developed for reducing the
bandwidth required for television pictures~\cite{rle}, and later used in
image compression and fax formats, among other things.

We will define a data type for storing run-length encoded lists, and use the
type system to ensure that the encoding is sound with respect to the
original list. To begin, we define a function which constructs a list by
repeating an element a given number of times. We will need this for
explaining the relationship between compressed and uncompressed data.

\begin{lstlisting}
rep : Nat -> a -> List a
rep Z x = []
rep (S k) x = x :: rep k x
\end{lstlisting}

Using this, and the concatenation operator for \TC{List} (\FN{++}, which is
defined like \FN{append}), we can describe what it means to be a run-length
encoded list:

\begin{lstlisting}
data RunLength : List ty -> Type where
     Empty : RunLength []
     Run : (n : Nat) -> (x : ty) -> (rle : RunLength more) ->
           RunLength (rep (S n) x ++ more)
\end{lstlisting}

%We can read this as:

\begin{itemize}
\item \DC{Empty} is the run-length encoding of the empty list \DC{[]}
\item Given a length \VV{n}, an element \VV{x}, and an encoding \VV{rle} of the
list \VV{more}, \DC{Run} \VV{n} \VV{x} \VV{rle} is the encoding
of the list \texttt{rep (S n) x ++ more}. That is, $n+1$ copies of \texttt{x}
followed by \VV{more}.
\end{itemize}

We use \texttt{S n} to ensure that \texttt{Run} always increases the length
of the list, but otherwise we make no attempt (in the type) to ensure that
the encoding is optimal; we merely ensure that the encoding is sound with
respect to the encoded list.
Let us try to write a function which uncompresses a run-length encoded list,
beginning with a partial definition:

\begin{lstlisting}
uncompress : RunLength {ty} xs -> List ty
uncompress rle = ?uncompress_rhs
\end{lstlisting}

\textbf{Note:} The \VV{\{ty\}} syntax gives an explicit value for the implicit
argument to \TC{RunLength}. This is to state in the type that the element type
of the list being returned is the same as the element type in the encoded
list.

Like our initial implementation of \FN{length} on \TC{Vect}, we might be
tempted to return \VV{xs} directly, since the index of the encoded list gives
us the original uncompressed list. However, if we check the type of
\texttt{uncompress\_rhs}, we see that \texttt{xs} isn't available at run time:

\begin{lstlisting}
 0 xs : List ty
   rle : RunLength xs
------------------------------
uncompress_rhs : List ty
\end{lstlisting}

This is a good thing: if the uncompressed list were available at
run-time, there would have been no point in compressing it!
We can still take advantage of having the uncompressed list available
as part of the type, though, by refining the type of \FN{uncompress}:

\begin{lstlisting}
data Singleton : a -> Type where
     Val : (x : a) -> Singleton x

uncompress : RunLength xs -> Singleton xs
\end{lstlisting}

A value of type \texttt{Singleton x} has exactly one value, \texttt{Val x}.
The type of \FN{uncompress} expresses that the uncompressed list
is guaranteed to have the same value as the original list, although it must
still be reconstructed at run-time. We can implement it as follows:

\begin{lstlisting}
uncompress : RunLength xs -> Singleton xs
uncompress Empty = Val []
uncompress (Run n x y) = let Val ys = uncompress y in
                             Val (x :: (rep n x ++ ys))
\end{lstlisting}

\textbf{Aside:} This implementation was generated by type-directed program
synthesis~\cite{polikarpova_program_2016}, rather than written by hand,
taking advantage of the explicit relationship given in the type between the
input and the output. The definition of \TC{RunLength} more or less directly
gives the uncompression algorithm, so we should not need to write it again!

The \texttt{0} multiplicity, therefore, allows us to reason about values
at compile time, without requiring them to be present at run time.
Furthermore, QTT gives us a guarantee of erasure, as well as an
explicit type level choice as to whether an index is erased or not.

\subsection{Linearity}

\label{sec:linearity}

An argument with multiplicity \texttt{0} is guaranteed to be erased at run
time. Correspondingly, an argument with multiplicity \texttt{1} is guaranteed
to be used exactly once. The intuition, similar to that of Linear
Haskell~\cite{linhaskell}, is that, given a function type of the form\ldots

\begin{lstlisting}
f : (1 x : a) -> b
\end{lstlisting}

\ldots then, if an expression \texttt{f x} is evaluated exactly once, 
\texttt{x} is evaluated exactly once in the process. 
QTT is a new core language, and the combination of linearity with dependent
types has not yet been extensively explored. Thus, we consider
the multiplicity \texttt{1} to be experimental, and in general Idris 2
programmers can get by without it---nothing in the Prelude exposes an interface
which requires it. Nevertheless, we have found
that an important application of linearity is in controlling resource
usage.
In the rest of this section, we describe two examples of this. First, 
we show in general how linearity can prevent us duplicating an argument, which can be
important if the argument represents an external resource; then, we give a 
more concrete example showing how the \TC{IO} type described in
Section~\ref{sec:interactive} is implemented.

\subsubsection{Example 1: Preventing Duplication}

To illustrate multiplicity \texttt{1} in practice, we can try (and fail!) to
write a function which duplicates a value declared as ``use once'',
interactively:

\begin{lstlisting}
dup : (1 x : a) -> (a, a)
dup x = ?dup_rhs
\end{lstlisting}

Inspecting the \texttt{dup\_rhs} hole shows that we have:

\begin{lstlisting}
 0 a : Type
 1 x : a
-------------------------------------
dup_rhs : (a, a)
\end{lstlisting}

So, \texttt{a} is not available at run-time, and \texttt{x} must be used
exactly once in the definition of \texttt{dup\_rhs}. We can write a
partial definition:

\begin{lstlisting}
dup x = (x, ?second_x)
\end{lstlisting}

However, if we check the hole \texttt{second\_x} we see that \texttt{x}
is not available, because there was only 1 and it has already been consumed:

\begin{lstlisting}
 0 a : Type
 0 x : a
-------------------------------------
second_x : a
\end{lstlisting}

We see the same result if we try \texttt{dup x = (?second\_x, x)}. If
we persist, and try\ldots

\begin{lstlisting}
dup x = (x, x)
\end{lstlisting}

\ldots then Idris reports ``\texttt{There are 2 uses of linear name x}''.

\textbf{Remark:} As we noted earlier, only usages in the body of a definition
count. This means we can still parameterise data by linear variables. For
example, if we have an \texttt{Ordered} predicate on lists, we can write
an \FN{insert} function on ordered linear lists:

\begin{lstlisting}
insert : a -> (1 xs: List a) -> (0 _ : Ordered xs) -> List a
\end{lstlisting}

The use in \TC{Ordered} does not count, and since \TC{Ordered} has multiplicity
$0$ it is erased at run time, so any occurrence of \texttt{xs} when
building the \TC{Ordered} proof also does not count.

\subsubsection{Example 2: I/O in Idris 2}

\label{sec:io}

Like Idris 1 and Haskell, Idris 2 uses a parameterised type \texttt{IO} to
describe interactive actions. Unlike the previous implementation, this is
implemented via a function which takes an abstract representation of the
outside world, of primitive type \texttt{\%World}:

\begin{lstlisting}
PrimIO : Type -> Type
PrimIO a = (1 x : %World) -> IORes a
\end{lstlisting}

The \texttt{\%World} is consumed exactly once, so
it is not possible to use previous worlds (after all, you
can't unplay a sound, or unsend a network message). It returns an
\texttt{IORes}:

\begin{lstlisting}
data IORes : Type -> Type where
     MkIORes : (result : a) -> (1 w : %World) -> IORes a
\end{lstlisting}

This is a pair of a result (with usage $\omega$), 
and an updated world state. The
intuition for multiplicities in data constructors is the same as 
in functions: here, if \texttt{MkIORes result w} is evaluated exactly once, then
the world \texttt{w} is evaluated exactly once.
We can now define \texttt{IO}:

\begin{lstlisting}
data IO : Type -> Type where
     MkIO : (1 fn : PrimIO a) -> IO a
\end{lstlisting}

There is a primitive \texttt{io\_bind} operator (which we can use to implement
$\mathtt{>>=}$), which guarantees that an action and its continuation are
executed exactly once:

\label{sec:iobind}
\begin{lstlisting}
io_bind : (1 act : IO a) -> (1 k : a -> IO b) -> IO b
io_bind (MkIO fn) = \k => MkIO (\w => let MkIORes x' w' = fn w
                                          MkIO res = k x' in res w')
\end{lstlisting}

The multiplicities of the \texttt{let} bindings are inferred from the values
being bound. Since \texttt{fn w} uses \texttt{w}, which is required to be
linear from the type of \texttt{MkIO}, \texttt{MkIORes x' w'} must itself be
linear, meaning that \texttt{w'} must also be linear. It can be informative
to insert a hole:
%to see how the multiplicities are updated:

\begin{lstlisting}
io_bind (MkIO fn)
    = \k => MkIO (\w => let MkIORes x' w' = fn w in ?io_bind_rhs)
\end{lstlisting}

This shows that, at the point \texttt{io\_bind\_rhs} is evaluated, we have
consumed \texttt{fn} and \texttt{w}, and we still have to run the continuation
\texttt{k} once, and use the updated world \texttt{w'} once:

\begin{lstlisting}
 0 b : Type
 0 a : Type
 0 fn : (1 x : %World) -> IORes a
 1 k : a -> IO b
 0 w : %World
 1 w' : %World
   x' : a
-------------------------------------
io_bind_rhs : IORes b
\end{lstlisting}

This implementation of \texttt{IO} is similar to the Haskell
approach~\cite{awkward}, with two differences:

\begin{enumerate}
\item The \texttt{\%World} token is guaranteed to be consumed exactly once, so
there is a type level guarantee that the outside world is never duplicated or
thrown away.
\item There is no built-in mechanism for exception handling, because the
type of \texttt{io\_bind} requires that the continuation is executed
exactly once. So, in \texttt{IO} primitives, we must be explicit about
where errors can occur. One can, however, implement higher level abstractions
which allow exceptions if required.
\end{enumerate}

Linearity is, therefore, fundamental to the implementation of \TC{IO} in
Idris 2. Fortunately, none of the implementation details need to be exposed
to application programmers who are using \TC{IO}.
However, once a programmer has an understanding of the combination of linear
and dependent types, they can use it to encode and verify more sophisticated
APIs. %, as we will see in the next section.

\section{Linear Resource Usage Protocols}

\label{sec:protocols}

The \texttt{IO} type uses a
linear resource representing the current state of the outside world. But,
often, we need to work with other resources, such as files, network sockets,
or communication channels. In this section, we introduce an extension of
\texttt{IO} which allows creating and updating linear resources, and show how
to use it to implement a resource usage protocol for an automated teller
machine (ATM).

\subsection{Initialising Linear Values}

In QTT, multiplicities are associated with \emph{binders}, not with return
values or types. This is a design decision of QTT, rather than of Idris, and has
the advantage that we can use a type linearly or not, depending on
context. We can write functions that create values to be used linearly by
using \emph{continuations}, for example, to create a new linear array:

\begin{lstlisting}
newArray : (size : Int) -> (1 k : (1 arr : Array t) -> a) -> a
\end{lstlisting}

The array must be used exactly once in the scope of \texttt{k}, and if this is
the only way of constructing an \texttt{Array}, then all arrays are guaranteed
to be used linearly, so we can have in-place update. As a matter of taste, however, we may not want to write
programs with explicit continuations. Fortunately, \texttt{do} notation can
help; recall the bind operator for \TC{IO}:

\begin{lstlisting}
(>>=) : IO a -> (a -> IO b) -> IO b
\end{lstlisting}

This allows us to chain an IO action and its continuation, and \texttt{do}
notation gives syntactic sugar for translating into applications of
$\mathtt{>>=}$. Therefore, we can define an alternative $\mathtt{>>=}$
operator for chaining actions which return linear values. We will achieve
this by defining a new type for capturing interactive actions, extending
\texttt{IO}, and defining its bind operator.

\subsection{Linear Interactive Programs}

First, we define how many times the result of an operation can be used.
These correspond to the multiplicities $0$, $1$ and $\omega$:

\begin{lstlisting}
data Usage = None | Linear | Unrestricted
\end{lstlisting}

We declare a data type \TC{L}, which describes interactive programs that
produce a result with a specific multiplicity. We choose a short name \TC{L}
since we expect this to be used often:

\begin{lstlisting}
data L : {default Unrestricted use : Usage} ->
         Type -> Type where
\end{lstlisting}

In Section~\ref{sec:implicits} we described implicit arguments, which are
solved by unification. Here, \VV{use} is an implicit argument, and the
\texttt{default Unrestricted} annotation means that if its value cannot be
solved by unification, it should have a default value of \texttt{Unrestricted}.

\textbf{Aside:} \TC{L} is defined in a library \TC{Control.Linear.LIO},
distributed with Idris 2. In fact, its type is more general;
\texttt{L : (io : Type -> Type) -> \{use : Usage\} -> Type -> Type}. This
allows us to extend \emph{any} monad with linearity, not just \texttt{IO},
but this generality is not necessary for the examples in this paper.

Like \TC{IO}, \TC{L} provides the operators \TC{pure} and $\mathtt{>>=}$.
However, unlike \TC{IO}, they need to account for variable usage. One limitation
of QTT is that it does not yet support quantity polymorphism, so we must
provide separate \TC{pure} operators for each quantity:

\begin{lstlisting}
pure  : (  x : a) -> L         a
pure0 : (0 x : a) -> L {use=0} a
pure1 : (1 x : a) -> L {use=1} a
\end{lstlisting}

Idris translates integer literals using the \TC{fromInteger} function. We
have defined a \TC{fromInteger} function that maps \texttt{0} to
\DC{None} and \texttt{1} to \DC{Linear} which allows us to use integer
literals as the values for the \VV{use} argument.

The type of $\mathtt{>>=}$ is more challenging. In order to take advantage of
\texttt{do}-notation, we need a single $\mathtt{>>=}$ operator for chaining
an action and a continuation, but there
are several possible combinators of variable usage. Consider:

\begin{itemize}
\item The action might return an erased, linear or unrestricted value.
\item Correspondingly, the continuation must bind its argument at multiplicity
$0$, $1$ or $\omega$.
\end{itemize}

In other words, the type of the continuation to $\mathtt{>>=}$ depends on the
usage of the result of the action. We can therefore take advantage of
first-class types and \emph{calculate} the continuation type. Given the
usage of the action (\VV{u\_a}), the usage of the continuation (\VV{u\_k})
and the return types of each, \VV{a} and \VV{b}, we calculate:

\begin{lstlisting}
ContType : (u_a : Usage) -> (u_k : Usage) -> Type -> Type -> Type
ContType None u_k a b = (0 _ : a) -> L {use=u_k} b
ContType Linear u_k a b = (1 _ : a) -> L {use=u_k} b
ContType Unrestricted u_k a b = a -> L {use=u_k} b
\end{lstlisting}

Then, we can write a type for $\mathtt{>>=}$ as follows:

\begin{lstlisting}
(>>=) : {u_a : _} ->
        (1 _ : L {use=u_a} a) ->
        (1 _ : ContType u_act u_k a b) -> L {use=u_k} b
\end{lstlisting}

The continuation type is calculated from the usage of the first action, and is
correspondingly needed in the implementation, so \VV{u\_a} is run time
relevant. However, in practice it is removed by inlining.
Fortunately, the user of \TC{L} need not worry about these details. They
can freely use \texttt{do}-notation and let the type checker take care of
variable usage for them.

%, taking the type of $\>\>=$ to be one of:

%\begin{lstlisting}
%(>>=) : (1 _ : L {use=0} a) ->
%        (1 _ : (0 _ : a) -> L {use} b) -> L {use} b
%(>>=) : (1 _ : L {use=1} a) ->
%        (1 _ : (1 _ : a) -> L {use} b) -> L {use} b
%(>>=) : (1 _ : L a) ->
%        (1 _ : a -> L {use} b) -> L {use} b
%\end{lstlisting}

Finally, for developing linear wrappers for
\TC{IO} libraries, we allow lifting \TC{IO} actions:

\begin{lstlisting}
action : IO a -> L a
\end{lstlisting}

We use \FN{action} for constructing primitives. Note that we will not be
able to bypass any linearity checks this way, since it does not promise to
use the \TC{IO} action linearly, so we cannot pass any linear resources to
an \FN{action}. The implementation of \TC{L} is via a well-typed
interpreter~\cite{augustsson_exercise_1999}, a standard pattern in
dependently typed programming.

%We will return to the implementation of \TC{L} shortly, in
%Section~\ref{sec:implementl}. First, let us look at a small example of using
%\TC{L} to implement a state machine, with the transitions checked at
%compile time.

\subsection{Example: An ATM State Machine}

We can use linear types to encode the state of a resource, and implement
operations in \TC{L} to ensure that they are only executed when the resource is
in the appropriate state.  For example, an ATM should only dispense cash when a
user has inserted their card and entered a correct PIN. This is a typical
sequence of operations on an ATM:

\begin{itemize}
\item A user inserts their bank card.
\item The machine prompts the user for their PIN, to check the user is
entitled to use the card.
\item If PIN entry is successful, the machine prompts the user for an amount of
money, and then dispenses cash.
\end{itemize}

\begin{figure}[h]
\begin{center}
\includegraphics[width=10cm]{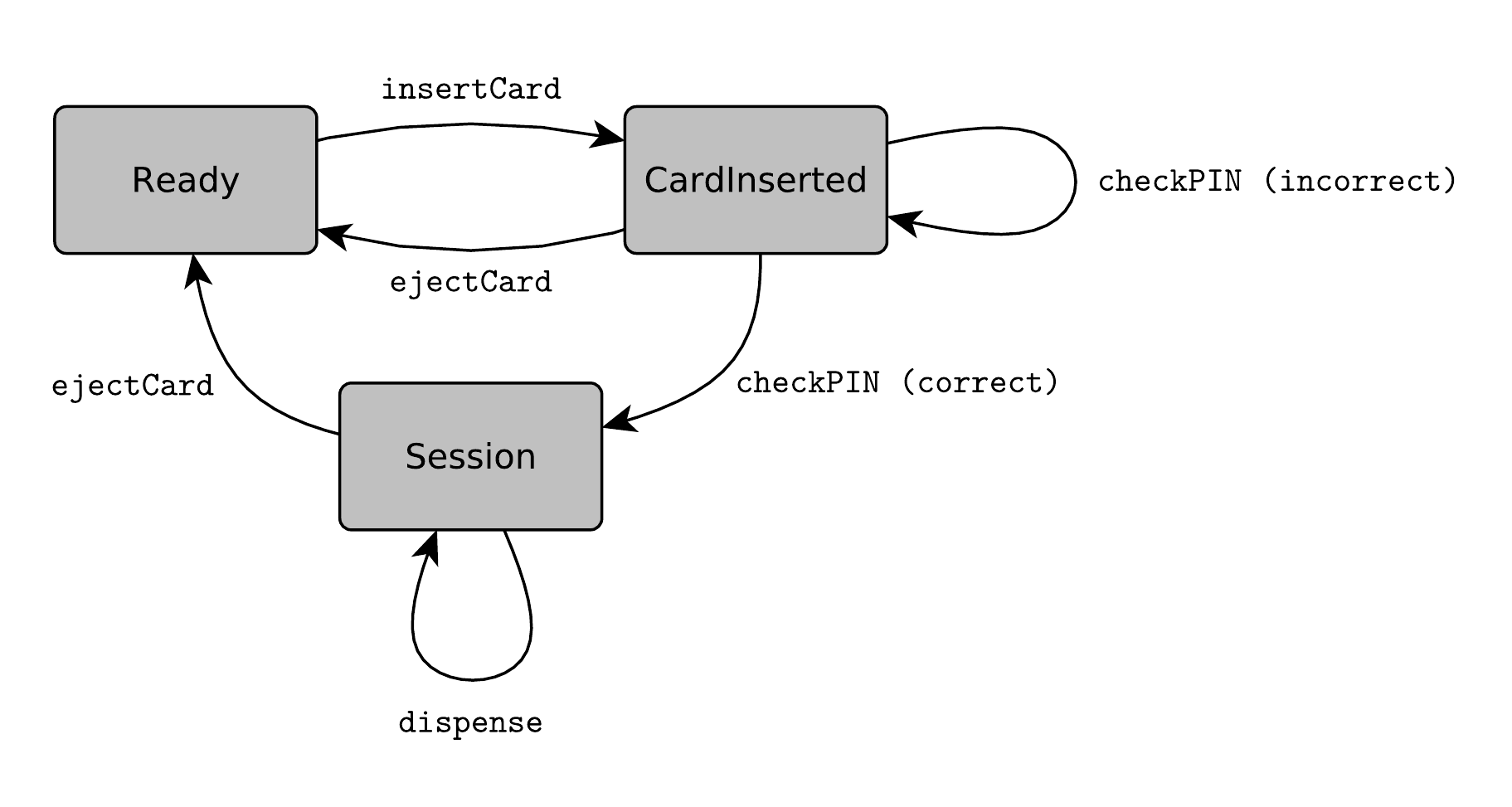}
\end{center}
\caption{A state machine describing the states and operations on an ATM}
\label{fig:atm}
\end{figure}

Figure \ref{fig:atm} defines, at a high level, the states and operations on
an ATM, showing when each operation is valid. We will define these operations
as functions in the \TC{L} type, using a linear reference to a representation
of an ATM. We define the valid states of the ATM as a data type, then an
ATM type which is parameterised by its current state, which is one of:

\begin{itemize}
\item \DC{Ready}: the ATM is ready and waiting for a card to be inserted.
\item \DC{CardInserted}: there is a card inside the ATM but the PIN entry
is not yet verified.
\item \DC{Session}: there is a card inside the ATM and the PIN has been
verified.
\end{itemize}

\begin{lstlisting}
data ATMState = Ready | CardInserted | Session
data ATM : ATMState -> Type
\end{lstlisting}

We leave the definition of \TC{ATM} abstract. In practice, this is where we
would need to handle implementation details such as how to access and update a
user's bank account. For the purposes of this example, we are only interested
in encoding the high level state transitions in types.
We will need functions to initialise and shut down the reference:

\begin{lstlisting}
initATM  : L {use=1} (ATM Ready)
shutDown : (1 _ : ATM Ready) -> L ()
\end{lstlisting}

\begin{lstlisting}[caption={Operations on an ATM},label=atmops,float=h]
data HasCard : ATMState -> Type where
     HasCardPINNotChecked : HasCard CardInserted
     HasCardPINChecked : HasCard Session

data PINCheck = CorrectPIN | IncorrectPIN

insertCard : (1 _ : ATM Ready) -> L {use=1} (ATM CardInserted)
checkPIN : (1 _ : ATM CardInserted) -> (pin : Int) ->
           L {use=1}
             (Res PINCheck 
                  (\res => ATM (case res of
                                     CorrectPIN => Session
                                     IncorrectPIN => CardInserted)))
dispense : (1 _ : ATM Session) -> L {use=1} (ATM Session)
getInput : HasCard st => (1 _ : ATM st) ->
                         L {use=1} (Res String (const (ATM st)))
ejectCard : HasCard st => (1 _ : ATM st) -> L {use=1} (ATM Ready)
message : (1 _ : ATM st) -> String -> L {use=1} (ATM st)
\end{lstlisting}

\FN{initATM} creates a linear reference to an ATM in the initial state,
\TC{Ready}, which must be used exactly once. Correspondingly,
\FN{shutDown} deletes the linear reference.
Listing \ref{atmops} presents the types of the remaining operations, 
implementing the transitions from Figure \ref{fig:atm}.

Several details and design choices, are worthy of note. We have:
user-directed state transitions, where the
\emph{programmer} is in control over whether an operation succeeds;
general purpose operations, which do not change the state, and are part of
the machine's user interface; 
and,
machine-directed state transitions, where the \emph{machine} is in control
over whether an operation succeeds, for example the machine decides if PIN
entry was correct.

%. For example, only the machine knows,
%at run time, whether a PIN was entered successfully or not: this is the
%only way to enter the \DC{Session} state.

\paragraph*{User-directed state transitions}

The \FN{insertCard} card function takes a machine in the \DC{Ready} state,
and returns a new machine in the \DC{CardInserted} state. The type ensures
that we can only run the function with the machine in the appropriate state.
For \FN{dispense}, we need to satisfy the security property that the machine
can only dispense money in a validated session. Thus, it has an input state
of \DC{Session}, and the session remains valid afterwards.

For \FN{ejectCard}, it is only valid to eject the card if there is already
a card in the machine. This is true in \emph{two} states: \DC{CardInserted}
and \DC{Session}. Therefore, we define a \emph{predicate} on states which
holds for states where there is a card in the machine:

\begin{lstlisting}
data HasCard : ATMState -> Type where
     HasCardPINNotChecked : HasCard CardInserted
     HasCardPINChecked : HasCard Session
\end{lstlisting}

The type of \FN{ejectCard} then takes an input of type \TC{HasCard st},
which is a proof that the predicate holds for the machine's input state
\VV{st}.

\begin{lstlisting}
ejectCard : HasCard st => (1 _ : ATM st) -> L {use=1} (ATM Ready)
\end{lstlisting}

The notation \texttt{HasCard st => ...}, with the \texttt{=>} operator,
means that this is an \emph{auto implicit} argument. Like implicits,
and \texttt{default} implicits, these can be omitted. The type checker will
attempt to fill in a value by searching the possible constructors. In this
case, if \VV{st} is \DC{CardInserted}, the value will be
\DC{HasCardPINNotCheck}, and if it is \DC{Session}, the value will be
\DC{HasCardPINChecked}. Otherwise, Idris will not be able to find a value, and
will report an error.
Auto implicits are also used for interfaces,
corresponding to type classes in Haskell, although we do not use interfaces
elsewhere in this paper.

\paragraph*{General purpose operations}

The \FN{message} function displays a message to the user. Its type means that
we can display a message no matter the machine's state. Nevertheless, since an
ATM is linear, we must return a new reference. The \FN{getInput} function
reads input from the user, using the machine's keypad, provided that there
is a card in the machine. Again, this needs to return a new reference, along
with the input.
\TC{Res} is a \emph{dependent pair} type, where the first item is
unrestricted, and the second item is linear with a type computed from the
value of the first element. We describe this below in the context of
\FN{checkPIN}.

\paragraph*{Machine-directed state transitions}

The most interesting case is \FN{checkPIN}. In the other functions, the
programmer is in control of when state transitions happen, but in this case,
the transition may or may not succeed, depending on whether the PIN is
correct. To capture this possibility, we return the result in a dependent
pair type, \TC{Res}, defined as follows in the Idris Prelude:

\begin{lstlisting}
data Res : (a : Type) -> (a -> Type) -> Type where
     (#) : (val : a) -> (1 r : t val) -> Res a t
\end{lstlisting}

This pairs a value, \VV{val}, with a linear resource whose type is computed
from the value. This can be illustrated with a partial ATM program:

\begin{lstlisting}
runATM : L ()
runATM = do m <- initATM
            m <- insertCard m
            ok # m <- checkPIN m 1234
            ?whatnow
\end{lstlisting}

This program initialises an ATM, inserts a card, then checks whether the card
has the PIN 1234. Checking the PIN returns a \TC{Res}, which we deconstruct,
then we can inspect the hole \texttt{?whatnow} to see where to go from here:

\begin{lstlisting}
   ok : PINCheck
 1 m : ATM (case ok of { CorrectPIN => Session
                         IncorrectPIN => CardInserted })
------------------------------
whatnow : L ()
\end{lstlisting}

So, we have the result of the \TC{PINCheck}, and an updated \TC{ATM}, but
we will only know the state of the \TC{ATM}, and hence be able to make
progress in the protocol, if we actually check the returned result!
We cannot know statically what the next state of the machine is
going to be, but using first class types, we \emph{can} statically check that the
necessary dynamic check is made.
We could also use a sum type, such as \TC{Either}, for the
result of the PIN check, e.g. 

\begin{lstlisting}
checkPIN : (1 _ : ATM CardInserted) -> (pin : Int) ->
           L {use=1} (Either (ATM CardInserted) (ATM Session))
\end{lstlisting}

This would arguably be simpler. On the other hand, by returning an \texttt{ATM}
with an as yet unknown state, we can still run operations on the \texttt{ATM}
such as \FN{message} even before resolving the state. This might be useful
for diagnostics, or for user feedback, for example.
Listing~\ref{atmrun} shows one possible ATM protocol implementation, displaying
a message before checking the PIN, then dispensing cash if the PIN was valid.
Note that the protocol also requires the card to have been ejected before
the machine is shut down.

\begin{lstlisting}[caption={Example ATM protocol implementation},label=atmrun,float=h]
runATM : L ()
runATM = do m <- initATM
            m <- insertCard m
            ok # m <- checkPIN m 1234
            m <- message m "Checking PIN"
            case ok of
                 CorrectPIN => do m <- dispense m
                                  m <- ejectCard m
                                  shutDown m
                 IncorrectPIN => do m <- ejectCard m
                                    shutDown m
\end{lstlisting}

\textbf{Aside:} Linearity and exceptions do not mix well, since when we
catch an exception, we need to know what state the machine was in at the point
it was thrown in order to clean up effectively. On the other hand, if we have
to check every result as in Listing~\ref{atmrun}, we will end up with a lot
of nested \texttt{case} blocks, which are hard to read. As a compromise,
Idris provides a pattern matching bind notation~\cite{brady_tfp14}, which
allows us to code to a ``happy path'' and deal with alternatives as they
arise. For example:

\begin{lstlisting}
runATM : L ()
runATM = do m <- initATM
            m <- insertCard m
            CorrectPIN # m <- checkPIN m 1234
               | IncorrectPIN # m => ?failure
            ?success
\end{lstlisting}

The ``happy path'' is that the PIN was entered correctly. The alternative
we need to handle is the \DC{IncorrectPIN} case.

%\subsection{Implementing \TC{L}}

%\label{sec:implementl}

\section{Session Types via QTT}

\label{sec:sessions}

To illustrate how we can use quantities on a more
substantial example, let us consider how to use them to implement session
types. Session types~\cite{Honda1993,Honda2008} give types to communication channels, allowing us to
express exactly \emph{when} a message can be sent on a channel, ensuring that
communication protocols are implemented completely and correctly. There has
been extensive previous work on defining calculi for session types\footnote{A collection of implementations is available
at \url{http://groups.inf.ed.ac.uk/abcd/session-implementations.html}}. In
Idris 2, the combination of linear and dependent types means that we can
implement session types directly:

\begin{itemize}
\item \textbf{Linearity} means that a channel can only be accessed once, 
and once a message has been sent or received on a channel, the channel is
in a new state.
\item \textbf{Dependent Types} give us a way of describing protocols at the
type level, where progress on a channel can change according to values sent
on the channel.
\end{itemize}

A complete implementation of session types would be
a paper in itself, so we limit ourselves to dyadic session types in concurrent
communicating processes.  We assume that functions are \emph{total}, so
processes will not terminate early and communication will always
succeed. In a full library, dealing with \emph{distributed} as
well as \emph{concurrent} processes, we would also need to consider failures
such as timeouts and badly formed messages~\cite{Fowler2019}.

The key idea is to parameterise channels by the actions which will be
executed on the channel---that is, the messages which will be
sent and received---and to use channels linearly. We declare a
\texttt{Channel} type as follows:

\begin{code}
data Actions : Type where
     Send : (a : Type) -> (a -> Actions) -> Actions
     Recv : (a : Type) -> (a -> Actions) -> Actions
     Close : Actions

data Channel : Actions -> Type
\end{code}

Internally, \texttt{Channel} contains a message queue for bidirectional
communication.
Listing \ref{sessioniface} shows the types of functions for
initiating sessions, and sending and receiving messages.
In the type of \texttt{send}, we see that to send a value of type
\texttt{ty} we must have a channel in the state \texttt{Send ty next},
where \texttt{next} is a function that computes the rest of the protocol.
The type of \texttt{recv} shows that we compute the rest of the protocol
by inspecting the value received. We initiate concurrent sessions with
\texttt{fork}, and will discuss the details of this shortly.

\begin{code}[float=h,frame=single,caption={Initiating and executing
concurrent sessions},label=sessioniface]
send : (1 chan : Channel (Send ty next)) -> (val : ty) ->
       L {use=1} (Channel (next val))
recv : (1 chan : Channel (Recv ty next)) ->
       L {use=1} (Res ty (\val => Channel (next val)))
close : (1 chan : Channel Close) -> L ()
fork : ((1 chan : Server p) -> L ()) -> L {use=1} (Client p)
\end{code}

First, let us see how to describe dyadic protocols such that a \emph{client}
and \emph{server} are guaranteed to be synchronised. We describe protocols
via a \emph{global} session type:

\begin{code}
data Protocol : Type -> Type where
     Request : (a : Type) -> Protocol a
     Respond : (a : Type) -> Protocol a
     (>>=) : Protocol a -> (a -> Protocol b) -> Protocol b
     Done : Protocol ()
\end{code}

A protocol involves a sequence of \texttt{Request}s from a client to
a server, and \texttt{Response}s from the server back to the client.
For example, we could define a protocol (Listing~\ref{utilsproto})
in which a client sends a \texttt{Command} to either \texttt{Add} a pair of
\texttt{Int}s or \texttt{Reverse} a \texttt{String}.

\begin{code}[float=h,frame=single,caption={A global session type describing
a protocol where a client can request either adding two \texttt{Int}s or
reversing a \texttt{String}},label=utilsproto]
data Command = Add | Reverse

Utils : Protocol ()
Utils = do cmd <- Request Command
           case cmd of
                Add => do Request (Int, Int)
                          Respond Int
                          Done
                Reverse => do Request String
                              Respond String
                              Done
\end{code}

\texttt{Protocol} is a DSL for describing communication
patterns. Embedding it in a dependently typed host language gives us
dependent session types for free. We use the embedding to our advantage 
in a small way, by having the protocol depend on \texttt{cmd}, the command
sent by the client.
We can write functions to calculate the protocol for the client and the
server: 

\begin{code}
AsClient, AsServer : Protocol a -> Actions
\end{code}

We omit the definitions, but each translates \texttt{Request} and
\texttt{Response} directly to the appropriate \texttt{Send} or
\texttt{Receive} action.
We can see how \texttt{Utils} translates into a type for the client side by
running \texttt{AsClient Utils}:

\begin{code}
Send Command (\res => (ClientK 
    (case res of 
          Add => Request (Int, Int) >>= \_ => 
                 Respond Int >>= \_ Done
          Reverse => Request String >>= \_ => 
                     Respond String >>= \_ Done)
\end{code}

Most importantly, this shows us that the first client side operation must be
to send a \texttt{Command}. The rest of the type is calculated from the
command which is sent; \texttt{ClientK} is internal to \texttt{AsClient} and
calculates the continuation of the type. Using these, we can
define the type for \texttt{fork}.

\begin{code}
Client, Server : Protocol a -> Type
Client p = Channel (AsClient p)
Server p = Channel (AsServer p)

fork : ((1 chan : Server p) -> L ()) -> L {use=1} (Client p)
\end{code}

The type of \texttt{fork} ensures that the client and the server are working
to the same protocol, by calculating the channel type of each from the same
global protocol. Since each channel is linear, both ends of the protocol
must be run to completion.

\begin{code}[float=h,frame=single,caption={An implementation of a server
for the \texttt{Utils} protocol},label=utilserver]
utilServer : (1 chan : Server Utils) -> L ()
utilServer chan
    = do cmd # chan <- recv chan
         case cmd of
              Add => do (x, y) # chan <- recv chan
                        chan <- send chan (x + y)
                        close chan
              Reverse => do str # chan <- recv chan
                            chan <- send chan (reverse str)
                            close chan
\end{code}

Listing~\ref{utilserver} shows a complete implementation of a server for
the \texttt{Utils} protocol. However, we do not typically write a complete
implementation in one go. Idris 2's support for \emph{holes} means that it
is more convenient to write the server incrementally, in a type-driven
way. We begin with just a skeleton definition, and look at the hole for the
right hand side:

\begin{code}
utilServer : (1 chan : Server Utils) -> L ()
utilServer chan = ?utilServer_rhs

 1 chan : Channel (Recv Command (\res => ... ))
-------------------------------------
utilServer_rhs : L ()
\end{code}

Therefore, the first action on \texttt{chan} must be to receive a
\texttt{Command}:

\begin{code}
utilServer : (1 chan : Server Utils) -> L ()
utilServer chan
    = do cmd # chan <- recv chan
         ?utilServer_rhs

   cmd : Command
 1 chan : Channel (ServerK (case cmd of ...) ...)
------------------------------
utilServer_rhs : L ()                                       
\end{code}

We elide the full details of the type of \texttt{chan} at this stage, but at
the top level it suggests that we can make progress by a \texttt{case} split on
\texttt{cmd}:

\begin{code}
utilServer : (1 chan : Server Utils) -> L ()
utilServer chan
    = do cmd # chan <- recv chan
         case cmd of
              Add => ?process_add
              Reverse => ?process_reverse
\end{code}

We make essential use of dependent \texttt{case} here, in that both
branches have a different type which is computed from the value of the
scrutinee \texttt{cmd}, similarly to \FN{PrintfType} in
Section~\ref{sec:printf}.
Now, for each of the holes \texttt{process\_add} and \texttt{process\_reverse}
we see more concretely how the protocol should proceed. e.g. for
\texttt{process\_add}, we see we have to receive a pair of \texttt{Int}s, then
send an \texttt{Int}:

\begin{code}
   1 chan : Channel (Recv (Int, Int) (\res => 
                    (Send Int (\res => Close))))
     cmd : Command
  -------------------------------------
  process_add : L () 
\end{code}

Developing the server in this way shows programming as a \emph{dialogue}
with the type checker. Rather than trying to work out the complete program,
with increasing frustration as the type checker rejects our attempts, we
write the program step by step, and ask the type checker for more information
on the variables in scope and the required result.

\subsubsection*{Extension: Sending Channels over Channels}

A useful extension is to allow a server to start up more
worker processes to handle client requests. This would require sending the
server's \TC{Channel} endpoint to the worker process. However, we cannot do
this with \TC{send} as it stands, because the value sent must be of
multiplicity $\omega$, and the \TC{Channel} is linear. One way to support
this would be to refine \TC{Protocol} to allow flagging messages as linear,
then add:

\begin{code}
send1 : (1 chan : Channel (Send1 ty next)) -> (1 val : ty) ->
        L {use=1} (Channel (next val))
\end{code}

This takes advantage of QTT's ability to parameterise types by linear variables
like \texttt{val} here. A worker protocol, using the \FN{Utils} protocol
above, could then be described as follows, where a server forks a new 
worker process and immediately sends it the communication \TC{Channel} for
the client:

\begin{code}
MakeWorker : Protocol ()
MakeWorker = do Request1 (Server Utils); Done
\end{code}

We leave full details of this implementation for future work. It is,
nevertheless, a minor adaptation of the session types library.

\section{Related Work}

\paragraph*{Substructural Types}

Linear types~\cite{wadler-linear} and other substructural type systems have
several applications, e.g. verifying unique access to
external resources~\cite{Ennals2004} and as a basis for session
types~\cite{Honda1993}.  These applications typically use domain specific type
systems, rather than the generality which would be given by full dependent
types.  There are also several implementations of linear or other substructural
type systems in functional
languages~\cite{Tov2011a,granule,cleanuniq,Morris2016}.
While these languages do not
have full dependent types, Granule~\cite{granule} allows many of the
same properties to be expressed with a sophisticated notion of graded types
which allows quantitative reasoning about resource usage, and work is in
progress to add dependent types to Granule~\cite{granuleesop,granulepopl}. 
ATS~\cite{Shi2013} is a functional language with linear types with support for
theorem proving, which allows reasoning about resource usage and low level
programming. An important mainstream example of the benefit of substructural
type systems is Rust\footnote{\url{https://rust-lang.org/}}~\cite{rustbelt}
which guarantees memory safety of imperative programs without garbage
collection or any run time overhead, and is expressive enough to implement
session types~\cite{rustsession}.

Historically, combining linear types and dependent types in a fully
general way---with first-class types, and the full language available
at the type level---has been a difficult problem, primarily because it is not
clear whether to count variable usages in types. The problem can be
avoided~\cite{Krishnaswami-ldtp} by disallowing dependent linear functions or
by limiting the form of dependency~\cite{Gaboardi2013},
but these approaches limit expressivity. For
example, we may still want to reason about linear variables which have been
consumed. Or, as we saw at the end of Section~\ref{sec:sessions}, we may
want to use a linear value as part of the computation of another type.
Quantitative Type Theory~\cite{qtt,nuttin}, allows full dependent
types with no restrictions on whether variables are used in types or terms, by
checking terms at a specific multiplicity.

\paragraph*{Erasure}

While linearity has clear potential benefits in allowing reasoning about
effects and resource usage, one of the main motivations for using QTT
is to give a clear semantics for erasure in the type system.
We distinguish \emph{erasure} from
\emph{relevance}, meaning that erased arguments are still relevant during
type-checking, but erased at run time.
Early approaches in Idris include the notion of ``forced arguments'' and
``collapsible data types''~\cite{Brady2005}, which give a predictable, if
not fully general, method for determining which arguments can be erased.
Idris 1 uses a whole program analysis~\cite{matus-draft}, partly inspired
by earlier work on Erasure Pure Type Systems~\cite{mishralinger-erasure}
to determine which arguments can be erased, which works well in practice but
doesn't allow a programmer to require specific arguments to be erased,
and means that separate compilation is difficult.
The problem of what to erase also exists in Haskell to some extent, even
without full dependent types, when implementing zero cost
coercions~\cite{rolehaskell}.
Our experience of the $0$ multiplicity of QTT so far is that it provides the 
cleanest solution to the erasure problem, although we no longer infer which
other arguments can be erased.

% \paragraph*{Interactive Editing and Program Synthesis}
% 
% We have briefly discussed how QTT improves support for program synthesis
% by taking usage restrictions into account in the search. Program synthesis
% in Idris has not yet been explored deeply, and existing work on
% type-driven program synthesis~\cite{Polikar2016},
% resource-guided program synthesis~\cite{Knoth19} and example-driven
% program synthesis~\cite{Osera15}
% will provide important insight into improving program search. Nevertheless,
% even a brute force search of a hint database has (anecdotally)
% proved remarkably effective for small search problems.
% 
% There is also a lot of scope for using quantitative types in interactive
% editing support. To make linear dependent types practically
% useful and accessible to application developers, good interactive tooling
% is essential. Recent work on front end tooling~\cite{Robert2018} and
% structural editing with typed holes~\cite{hazel} will influence
% Idris 2.

\paragraph*{Reasoning about Effects}

One of the motivations for using QTT beyond expressing erasure in types is
that it provides a core language which allows reasoning about resource
usage---and hence, reasoning about interactions with external libraries.
Previous work on reasoning about effects and resources with dependent types
has relied on indexed monads~\cite{Atkey2009,McBride2011} or embedded DSLs
for describing effects~\cite{brady-tfp14}. These are effective, but
generally difficult to compose; even if we can compose effects in a
single EDSL, it is hard to compose multiple EDSLs, especially when
parameterised with type information. Other successful approaches
such as Hoare Type Theory~\cite{Ynot2008} are sufficiently expressive, but
difficult to apply in everyday programming. Having linear types in the core
language means that tracking state changes, which we have previously had to
encode in a state-tracking monad, is now possible directly in the language. We
can compose multiple resources by using multiple linear arguments.
Combining dependent and linear types, along with protocol descriptions in
\TC{L}, gives us similar power to
Typestate~\cite{typestate_2009,typestate_2014}, in that we can use dependency
to capture the state of a value in its type, and linearity to ensure that it is
always used in a valid state. First-class types gives us additional flexibility:
we can reason about state changes which are only known at run-time,
such as state changes which depend on a run-time check of a PIN in an ATM.

% The \texttt{App} library provides similar expressivity to runners of algebraic 
% effects~\cite{ahmanrunners}, which provide a mathematical model of resource
% management, and, like our \texttt{(>>=)} operator, ensure that continuations
% are run at most once.
% While our approach using \texttt{App} is not as expressive as, say,
% algebraic effects~\cite{Plotkin2009,Lindley2017}, monad
% transformers~\cite{Liang1995} or separation logic, it is composable with these
% more expressive approaches in exactly the same way as \texttt{IO}. For example,
% \texttt{App} could be used at the bottom of a monad transformer stack, or 
% as a way of instantiating a program built on algebraic effects or free monads.

\paragraph*{Session Types}

In Section~\ref{sec:sessions} we gave an example of using QTT
to implement Dyadic Session Types~\cite{Honda1993}. In previous 
work~\cite{brady-lambda2017} Idris has been experimentally extended with
uniqueness types, to support verification of concurrent protocols. However,
this earlier system did not support erasure, and as implemented it was hard
to combine unique and non-unique references. Our experience with QTT is that
its approach to linearity, with multiplicities on the binders rather than on
the types, is much easier to combine with other non-linear programs.

Given linearity and dependent types, we can already have dependent session
types, where, for example, the progress of a session depends on a message
sent earlier. Thus, the embedding gives us label-dependent session
types~\cite{labelst} with no additional cost. Previous work in exploring value-dependent
sessions in a dependently typed language~\cite{janplaces2019} is directly
expressible using linearity in Idris 2.  
We have not yet explored further extensions to
session types, however, such as multiparty session types~\cite{Honda2008},
dealing with exceptions during protocol execution~\cite{Fowler2019} or
dealing with errors in transmission in distributed systems.

\section{Conclusions and Further Work}

Implementing Idris 2 with Quantitative Type Theory in the core has immediately
given us a lot more expressivity in types than Idris 1.  For most day to day
programming tasks, expressing erasure at the type level is the most valuable
user-visible new feature enabled by QTT, in that it is unambiguous which
function and data arguments will be erased at run time.  Erasure has been a
difficulty for dependently typed languages for decades and until recently has been
handled in partial and unsatisfying ways (e.g.~\cite{brady_inductive_2003}).
Quantitative Types, and related recent work~\cite{matus-draft}, are the most
satisfying so far, in that they give the programmer complete control over what
is erased at run time. In future, we may consider combining QTT with inference
for additional erasure~\cite{matus-draft}.

The $1$ multiplicity enables programming with full linear dependent types.
Therefore reasoning about resources, which previously required heavyweight
library implementations, is now possible directly, in pure functions. We have
also seen, briefly, that quantities give more information when inspecting the
types of holes. More expressive types, with interactive editing tools.  make
programming a \emph{dialogue} with the machine, rather than an exercise in
frustration when submitting complete (but wrong!) programs to the type checker.

We have often found full dependent types, where a type is a first class
language construct, to be extremely valuable in developing libraries with
expressive interfaces, even if the programs which use those libraries do not
use dependent types much. The \texttt{L} type for embedding linear protocols
is an example of this, in that it allows a programmer to express precisely
not only \emph{what} a function does, but also \emph{when} it is allowed to
do it. It is important that the type system remains accessible to programmers,
however. Dependent and linear types are powerful concepts, and without care in
library design, can be hard to use. However, they don't have to be: they are based on
concepts that programmers routinely understand and use, such as using a
variable once and making assumptions about the relationships between data. 
A challenge for language and tool designers
is to find the right syntax and the right feedback mechanisms, so that
powerful verification tools are within reach of all software developers.

%\subsection{Future work}

While we have already found many benefits of being able to express
quantities in types, we have only just begun exploring, and have encountered
some limitations in the theory which we hope to address, perhaps adapting
ideas from related work~\cite{granulepopl}. Most
importantly, we would like to express \emph{polymorphic}
quantities. This may, for example, help give an appropriate type to
$\mathtt{>>=}$ taking into account that some monads guarantee to execute
the continuation exactly once, but others need more flexibility. Similarly,
like Granule~\cite{granule}, we may find it useful to use quantities other
than \texttt{0} and \texttt{1}, and the theory behind QTT supports
this.

We have not discussed performance in this paper, but for an interactive
system it is vital, and will be a primary concern in the near future.
Following~\cite{kovacs-smalltt}, Idris 2 minimises substitution of unification
solutions. Initial results are promising: Idris 2 is now self-hosting, and
builds itself in around 90 seconds\footnote{Dell XPS 13 Laptop, running Ubuntu 18.03 LTS}.
We are using the interactive development tools, especially holes, in developing
Idris itself.

Finally, an important application of reasoning about linear resource usage
is in implementing communication and security protocols correctly. The
\TC{Protocol} type in Section~\ref{sec:sessions} provides a preliminary example
which demonstrates the possibilites, but realistically it will need to
handle timeouts, exceptions and more sophisticated protocols. Implementing
these protocols correctly is difficult and error prone, and errors lead
to damaging security problems\footnote{e.g. \url{https://www.imperialviolet.org/2014/02/22/applebug.html}}.
But in describing a session type, we have explained a protocol in detail,
and the machine calculates a lot of information about how the protocol
proceeds. We should not let the type checker keep this information to itself!
Thus, interactive programming of protocols based on linear resource usage
gives a foundation for secure programming.

\bibliography{idris-qtt.bib}

\end{document}